\documentstyle[12pt,aasms4]{article}
 
\begin{document}
\title{A multidimensional hydrodynamic code for structure evolution in
cosmology}
\author{Vicent Quilis
, Jos\'e M$^{\underline{\mbox{a}}}$. Ib\'a\~nez
\and Diego S\'aez}
\affil{Departament d'Astronomia i
Astrof\'{\i}sica, Universitat de Val\`encia,
E-46100 Burjassot, Val\`encia, Spain}
\authoremail{vicente.quilis@uv.es}
 
\begin{abstract}
A cosmological multidimensional
hydrodynamic code is described and tested. This code is based on modern
{\it high-resolution shock-capturing} techniques.
It can make use of a linear or a parabolic cell reconstruction as well
as an approximate Riemann solver.
The code
has been specifically
designed for cosmological applications.
Two tests including shocks have been considered:
the first one is  a standard shock tube
and the second test involves a  spherically symmetric shock.
Various additional cosmological tests are also
presented. In this way,
the performance of the code is proved.
The usefulness of the code is discussed;
in particular,
this powerful tool is expected to be useful in order
to study the evolution of the hot gas component located inside
nonsymmetric cosmological structures.
 
\end{abstract}
 
\keywords{hydrodynamics -- methods: numerical --
large-scale structure of Universe -- Cosmology}
 
\newpage

\section{Introduction}

In this paper, a new hydrodynamical code is presented and tested. This code is
the multidimensional extension of a previous one described by Quilis et
al.(1994); it has been designed with the essential aim of simulating the
evolution of a nonsymmetric three-dimensional (3D) distribution of gas. The
features of this distribution are similar to those of the hot rarified gas
located inside cosmological structures. Further applications of the
multidimensional code will be presented elsewhere.
 
Even if the numerical techniques, the physical principles and the equations
involved in a code are appropriate, the hydrodynamical code must be tested in
order to discover unexpected failures and possible limitations. Six tests have
been selected spanning a significant set of situations which  allow us to check
all the relevant ingredients of the code. The chosen tests correspond to six
problems with known analytical or numerical solutions. These solutions are
compared with those obtained using our multidimensional code. Results are
encouraging.
 
Two types of methods can be used in order to study structure formation in
Cosmology: Analytical and numerical methods. Among the analytical methods,
several Eulerian and Lagragian hydrodynamical approaches have been proposed. In
these approaches, the Universe is considered as a fluid (see Bertschinger
1991). Let us mention, as examples, the {\it Zel'dovich solution} (Zel'dovich
1970), the {\it Adhesion model} (Gurbatov et al. 1989), and the {\it frozen
flow approximation} (Matarrese et al. 1992). These approaches apply beyond the
linear regime, but they have limitations. Zel'dovich's 3D solution only works
properly up to the mildly nonlinear regime and both the frozen flow and the
adhesion models introduce fictitious forces in order to avoid caustic
formation. On account of these limitations, it seems that the use of other
techniques such as the numerical ones is appropriate. N-body simulations are
used in the pressureless case and very robust hydrodynamical codes have been
proposed for studying collisional matter. Let us mention the code developed by
Cen (1992), which uses {\it artificial viscosity}, and the smooth particle
hydrodynamic codes originally developed by Gingold \& Monaghan (1977) and Lucy
(1977), independently. This paper is a preliminary study, in which modern
numerical techniques are implemented in order to design a code for future
applications to structure formation in Cosmology.

We have built up a multidimensional cosmological hydrodynamical code based on
{\it modern high-resolution shock-capturing} (HRSC)  techniques. These HRSC
methods were specifically designed for solving hyperbolic systems of
conservation laws and have two main features: they are at least second order
accurate on the smooth part of the flow and they give well resolved
nonoscillatory discontinuities (LeVeque 1992). By construction HRSC schemes
avoid using numerical artifacts,  such as the artificial viscosity, in order to
smear shocks. With HRSC techniques strong shocks are sharply solved, typically,
in two or three numerical cells, and they are free of spurious oscillations due
to the Gibbs phenomenon. This last property could be of crucial importance in
3D calculations where the numerical grid is constrained for obvious technical
reasons and has a poor resolution.

In recent years, HRSC methods have been applied widely in many astrophysical
fields: interacting stellar winds (see, e.g., Mellema et al. 1991), type II
supernovae explosions (see, e.g., M\"uller 1994) , relativistic jets
(Mart\'{\i} et al., 1995), etc. More recently, some codes --with cosmological
applications-- based on HRSC techniques have been presented. A very recent
multidimensional cosmological hydro-code built up using these techniques is the
one described by Ryu et al. (1993). In Quilis et al. (1994) we analyzed the
main features of a one-dimensional code, and in the present paper we are
concerned with the multidimensional version of this code.
 
Hereafter, $t$ stands for the cosmological time, $t_0$ is the age of the
Universe, $a(t)$ is the scale factor of a flat background. $\dot{X}$ stands for
the derivative of the function $X$ with respect to the cosmological time.
Function $\dot{a}/a$ is denoted by $H$. Hubble constant is the present value of
$H$; its value in units of $100 \ Km \ s^{-1} \ Mpc^{-1}$ is $h$. Velocities
are given in units of the speed of light. $\rho$ and $\rho_{_{B}}$ stand for
mass density and  background mass density, respectively. The density contrast
is $\delta=(\rho-\rho_{_{B}})/\rho_{_{B}}$. The background is flat. $p$ and
$\epsilon$ stand for pressure and internal energy per unit mass, respectively.

The plan of this paper is as follows: In Section 2, our numerical code is
described. In Section 3, the results of several tests are shown. Finally, a
general discussion is presented in Section 4.

\section{Equations and numerical procedure}
 
In this Section we are going to write the basic equations governing the
evolution of cosmological inhomogeneities as a {\it hyperbolic system of
conservation laws}. The mathematical properties of this kind of system have
been well studied (see, e.g., Lax 1973). Numerical algorithms have been
specifically designed for solving these systems of partial differential
equations (see, e.g., Yee 1989 and LeVeque 1992).
 
\subsection{Evolution equations in conservation form}

For spatial scales which are small enough
(see below), cosmological
inhomogeneities evolve according to the following equations (Peebles 1980):
\begin{equation}
\frac{\partial \delta}{\partial t} + \frac{1}{a}\nabla\cdot (1+ \delta)
{\vec v} = 0
\end{equation}
\begin{equation}
\frac{\partial {\vec v}}{\partial t} + \frac{1}{a}
({\vec v}\,\cdot\, \nabla){\vec v}
+ H{\vec v} = - \frac{1}{\rho a}\nabla p - \frac{1}{a} \nabla \phi
\end{equation}
\begin{equation}
\frac{\partial E}{\partial t} + {1\over a} \nabla\cdot[(E + p) {\vec v}] =
-3H(E+p) -H\rho {\vec v}^2 - \frac{ \rho {\vec v}}{a} \nabla \phi
\end{equation}
\begin{equation}
\nabla^2\phi = \frac{3}{2} H^2 a^2 \delta
\end{equation}
where ${\vec x}$, ${\vec v}=a(t)\frac{d{\vec x}}{dt}= (v_x, v_y, v_z)$, $E$
and $\phi(t,{\vec x})$ are,
respectively, the Eulerian coordinates, the peculiar velocity,
the total energy $E= \rho \epsilon + {1\over 2} \rho v^2$
($v^2 = v_x^2 + v_y^2 + v_z^2$),
 and the peculiar
Newtonian gravitational potential. Pressure gradients and gravitational
forces are the responsible for this evolution.
 
This approach applies if the following conditions 
are satisfied: a) The size of
the inhomogeneity is much smaller than the causal horizon size; thus,
background curvature is negligible,
b) velocities are much smaller than the
speed of light, and c) no strong local 
gravitational fields are present. These
conditions make a relativistic approach  unnecessary .
 
In 3D case, Eqs (1), (2) and (3) can be written as follows:
\begin{equation}
\frac{\partial \delta }{\partial t} + \frac{\partial}{\partial x^{\alpha}}
[(\delta + 1) \frac{v_{\alpha}}{a}] = 0
\end{equation}
\begin{eqnarray}
\frac{\partial}{\partial t}[(\delta +1)v_{\alpha}]
 + \frac{\partial}{\partial x^{\beta}}
[\frac{v_{\alpha}v_{\beta}(\delta +1)}{a}  +
\frac{p}{a\rho_{_{B}}} \delta_{{\alpha}{\beta}}] =
 -\frac{(\delta +1)}{a} \frac{\partial \phi}{\partial x^{\alpha}}
-(\delta + 1)v_{\alpha}H
\end{eqnarray}
\begin{eqnarray}
\frac{\partial E }{\partial t}  + \frac{\partial}{\partial x^{\alpha}}
[\frac{(E+p)v_{\alpha}}{a}] = -3H(E+p)- (\delta + 1)\rho_{_{B}}H v^2
-\frac{(\delta +1)v_{\alpha}\rho_{_{B}}}{a}
\frac{\partial \phi}{\partial x^{\alpha}}
\end{eqnarray}
\noindent
where $\alpha, \beta = 1, 2, 3 \, $, and
$\delta_{{\alpha}{\beta}}$ is Kronecker's delta.
 
Poisson's equation (4) is an elliptic equation. Its solution depends on the
boundary conditions. This equation
is used --at each time step-- to compute the source
term $\nabla \phi$ which appears
in Eqs (2) and (3). In multidimensional cases,
Poisson's equation is solved by using the Fast Fourier Transform (FFT)
algorithm (see Section 3.2.2). In 1D cases,
Poisson's equation reduces to an ordinary differential equation and,
consequently, the FFT is not necessary.
 
An equation of state $p=p(\rho,\epsilon)$ closes the system.
The ideal gas equation of state $p=(\gamma
-1)\rho\epsilon$ has been used in
all applications in this paper.
 
Let us focus on the evolutionary part of the above system.
The evolution equations (5)-(7) can be written in the form:\\
\begin{equation}
\frac{\partial {\vec u}}{\partial t} + \frac{\partial{\vec f(\vec u)}}
{\partial x}
+\frac{\partial{\vec g(\vec u)}}{\partial y} +
\frac{\partial{\vec h(\vec u)}}{\partial z}= {\vec s(\vec u)}
\end{equation}
\noindent
where ${\vec u}: \Re \times \Re^3 \rightarrow \Re^5$
is the vector of {\it unknowns}:
\begin{equation}
{\vec u} = [ \delta , m_x,m_y,m_z, E] \ \ ,
\end{equation}
\noindent
the three {\it flux} \ functions
${\vec F}^{\alpha} \equiv \{{\vec f}, {\vec g},{\vec h}\}$
in the spatial directions $x, y, z$, respectively:
$\Re^5 \rightarrow \Re^5$, are defined by
\begin{eqnarray}
{\vec f(\vec u)} = \left[  \frac{m_x}{a} ,
 \frac{m_x^2}{(\delta + 1)a} + \frac{p}{a\rho_{_{B}}} ,
\frac{m_xm_y}{(\delta+1)a},
   \frac{m_xm_z}{(\delta+1)a},
\frac{(E+p)m_x}{a(\delta + 1)} \right]
\end{eqnarray}
\begin{eqnarray}
{\vec g(\vec u)} = \left[   \frac{m_y}{a} , \frac{m_xm_y}{(\delta+1)a},
 \frac{m_y^2}{(\delta + 1)a} + \frac{p}{a\rho_{_{B}}},
\frac{m_ym_z}{(\delta+1)a},
\frac{(E+p)m_y}{a(\delta + 1)}\right]
\end{eqnarray}
\begin{eqnarray}
{\vec h(\vec u)} = \left[   \frac{m_z}{a} ,
  \frac{m_xm_z}{(\delta+1)a},\frac{m_ym_z}{(\delta+1)a},
\frac{m_z^2}{(\delta + 1)a} + \frac{p}{a\rho_{_{B}}},
\frac{(E+p)m_z}{a
(\delta + 1)}\right]
\end{eqnarray}
\noindent
and the {\it sources}
${\vec s} : \Re^5 \rightarrow \Re^5$ are
\begin{eqnarray}
{\vec s(\vec u)} = \left[ 0 \right.&,-&\frac{(\delta +1 )}{a}
\frac{\partial\phi}{\partial x}- Hm_x ,
-\frac{(\delta +1 )}{a}
\frac{\partial\phi}{\partial y}- Hm_y ,
-\frac{(\delta +1 )}{a}
\frac{\partial\phi}{\partial z}- Hm_z,\nonumber \\
&-& \left. 3H(E+p)
- \frac{\rho_{_{B}}H m^2}{(\delta + 1)}
- \frac{m_x\rho_{_{B}}}{a}\frac{\partial\phi}{\partial x}
- \frac{m_y\rho_{_{B}}}{a}\frac{\partial\phi}{\partial y}
- \frac{m_z\rho_{_{B}}}{a}\frac{\partial\phi}{\partial z}\right]
\end{eqnarray}
where $m_x=(\delta +1)v_x$, $m_y=(\delta +1)v_y$, and $m_z=(\delta +1)v_z$.
 
System (8) is
a three-dimensional {\it hyperbolic system of conservation laws} with
sources ${\vec s}({\vec u})$. Let us introduce the Jacobian
matrices associated to the fluxes:
\begin{equation}
{\bf \cal A}^{\alpha} = \frac{\partial{\vec F}^{\alpha}({\vec u})}
{\partial {\vec u}}
\label{A}
\end{equation}
\noindent
hence, hyperbolicity demands that
any real linear combination of the Jacobian matrices
$\xi_{\alpha} {\bf \cal A}^{\alpha}$ should be diagonalizable with
real eigenvalues (LeVeque 1992). Knowledge of this fundamental
property is not only of theoretical interest but is also of
crucial importance from the numerical point of view (see below).
 
The spectral decompositions of the above Jacobian matrices in each direction,
i.e., the {\it eigenvalues} and {\it eigenvectors} are explicitly written in
the Appendix A.
For Newtonian dynamics they can be found in, e.g., Glaister (1988).
Let us point out that, in our case,
the expansion factor does not influence the
general features of the spectral decomposition derived in Newtonian dynamics.
 
The sources do not contain any term with differential operators acting on
hydrodynamical variables ${\vec u}$. This is an important property consistent
with the fact that the left hand side of Eq.\ (8) defines a hyperbolic system
of conservation laws.

\subsection{Numerical case}
 
The main features of our multidimensional algorithm are the following:
 
\begin{enumerate}
\item It is written in {\it conservation form}.
This is a very important property for a numerical algorithm designed
to solve, numerically, a hyperbolic system of conservation laws. That is,
in the absence of sources, those quantities
that ought to be conserved --according to the differential equations--
are exactly conserved in the difference form.
 
Let us comment how this fundamental idea is carried out in practice.
The integral form of system (8) is
\begin{equation}
\int_{\Omega} \frac {\partial {\vec u}} {\partial t} d\Omega +
\int_{\Omega} \frac {\partial {\vec F}^{\alpha}}{\partial x^{\alpha}} d\Omega
 = \int_{\Omega} {\vec s} d\Omega
\label{int}
\end{equation}
\noindent
where the domain $\Omega$
= $[t, t+\Delta t] \times [x^{\alpha}, x^{\alpha}+\Delta x^{\alpha}]$
$\forall \alpha$ defines a computational cell.
Gauss' theorem allows us to write the above system in the
following conservative form, well-adapted
to numerical applications:
\begin{eqnarray}
(\displaystyle{\langle{\vec u}_{t+\Delta t}\rangle} -
\displaystyle{\langle{\vec u}_{t}\rangle}){\Delta V} & = &
- \left( \int_{\Sigma_{x^1+\Delta x^1}} \widehat{{\vec f}} dt dx^2
dx^3 -
\int_{\Sigma_{x^1}} \widehat{{\vec f}} dt dx^2 dx^3 \right) \nonumber \\
&-& \left( \int_{\Sigma_{x^2+\Delta x^2}} \widehat{{\vec g}} dt dx^1 dx^3 -
\int_{\Sigma_{x^2}} \widehat{{\vec g}} dt dx^1 dx^3 \right)
\nonumber \\
&-& \left( \int_{\Sigma_{x^3+\Delta x^3}} \widehat{{\vec h}} dt dx^1 dx^2 -
\int_{\Sigma_{x^3}} \widehat{{\vec h}} dt dx^1 dx^2 \right)
+ \int_{\Omega} {\vec s} d\Omega
\label{integ}
\end{eqnarray}
\noindent
where
\begin{equation}
\displaystyle{\langle{\vec u}\rangle} = \frac{1}{\Delta V}
\int_{\Delta V} {\vec u} dx^1 dx^2 dx^3
\end{equation}
\begin{equation}
{\Delta V} = \int^{x^1+\Delta x^1}_{x^1}
\int^{x^2+\Delta x^2}_{x^2}
\int^{x^3+\Delta x^3}_{x^3}
dx^1 dx^2 dx^3
\end{equation}
 
Hence, the variation in time
of quantities ${\langle{\vec u}\rangle}$ within $\Omega$
--apart from the sources--
is given by the fluxes ${\vec F}^{\alpha}$ on its boundary $\partial \Omega$.
 
The symbol $(\,\,\widehat{}\,\,)$ appearing on the fluxes on the right-hand
side of (16) denote that they are calculated at
the cell interfaces between the corresponding numerical cells
by solving local Riemann problems
(i.e., initial value problems with discontinuous data).
 
The discretized version of (16) is obvious if we take into
account that, at each time step $t=t^{n}$, data
${\vec u}^{n}_{i,j,k}$
are the {\it cell-average} of the
variables ${\vec u}(x,y,z,t)$:
\begin{eqnarray}
{\vec u}^{n}_{i,j,k} =
\frac{\int_{x_{i-1/2}}^{x_{i+1/2}}\int_{y_{j-1/2}}^{y_{j+1/2}}
\int_{z_{k-1/2}}^{z_{k+1/2}}
\vec u \rm(x,y,z,t^{n}) dxdydz}{\Delta x_{i}\Delta y_{j}\Delta z_{k}} \ ,
\end{eqnarray}
\noindent
where the set of indices ($i,j,k$) stand for
the $x,y,z$ directions , respectively,
of a particular cell at a given
time level. Indices $l-1/2$ and $l+1/2$ (for $l=i,j,k$)
stand for the corresponding interfaces in each direction.
These cell-averaged quantities are evolved in time (see below).
 
\item {\it Reconstruction procedure}.
In order to increase spatial accuracy we have used
two cell-reconstruction techniques. We have implemented a linear
reconstruction, with
the {\it minmod} function (Quilis et al. 1994) as a slope limiter.
This limiter warranties that the method does not increase the total
variation (TV), i.e.,
\begin{eqnarray}
TV({\vec u}^{n+1}) \le TV({\vec u}^n)
\end{eqnarray}
\noindent
where the operation $TV$ is defined by
\begin{eqnarray}
TV({\vec u}^n) = \sum_{\forall l} \mid {\vec u}^n_{l+1} - {\vec u}^n_{l} \mid
\end{eqnarray}
\noindent
and subindex $l$ stands for a generic spatial direction.
 
With this reconstruction our algorithm is a MUSCL-version (from
monotonic upwind schemes for conservation laws, Van Leer 1979) and
second order accurate in space.
According to LeVeque (1992, page 184) MUSCL is a
TVD (total variation diminishing) for scalar conservation laws.
 
We have also implemented a parabolic reconstruction (PPM) subroutine according
to the procedure derived by Colella \& Woodward (1984). With the parabolic
reconstruction the algorithm is third order accurate in space. Statements on
the order of accuracy of an algorithm should be taken carefully when applied to
a system and, moreover, when the numerical solution develops discontinuities.
However, it can be proved that the above statements (in the sense of the local
truncation error) are correct, at least for scalar equations.
 
Hence, from the cell-averaged quantities ${\vec u}^{n}_{i,j,k}$ we construct,
in each direction, a piecewise linear or parabolic function which preserves
monotonicity. Thus, the quantities, ${\vec u}_{i+{1\over 2},j,k}^R, {\vec
u}_{i,j+{1\over 2},k}^R,{\vec u}_{i,j,k+{1\over 2}}^R$ and ${\vec u}_{i+{1\over
2},j,k}^L,{\vec u}_{i,j+{1\over 2},k}^L, {\vec u}_{i,j,k+{1\over 2}}^L$, can be
computed; the superindices $R$ and $L$ stand for the values at the both sides
of a given interface between neighbourhood cells. These values at each side of
a given interface allow us to define the local Riemann problems. The numerical
fluxes can be computed through the solution of these local Riemann problems.
 
\item {\it Numerical fluxes} at interfaces. We have used a linearized
Riemann solver following an approach similar to the one
described by Roe (1981). The procedure, in each direction,
starts by constructing the corresponding numerical flux according
to Roe's prescription and in order to do that it is necessary to know the
spectral decomposition
of the Jacobian matrix ${\cal A}^{\alpha}$. Let us, for the sake of
simplicity, focus on the $x$-direction. The numerical flux associated
to the $x$-direction is:
\begin{eqnarray}
{\widehat {{\vec f}}}_{i+{1\over 2},j,k}
 = \frac{1}{2}\left(
 {\vec f}({\vec u}_{i+{1\over 2},j,k}^{L}) +
{\vec f}({\vec u}_{i+{1\over 2},j,k}^{R})
- \sum_{{\eta} = 1}^{5} \mid
\widetilde{\lambda}_{\eta}^x\mid
\Delta \widetilde {\omega}_{\eta}
{\widetilde {\vec R}}^x_{\eta} \right)
\end{eqnarray}
\noindent
where
${\widetilde {\lambda}}_{\eta}^x$ and ${\widetilde {\vec R}}_{\eta}^x$
($\eta=1,2,3,4,5$) are, respectively, the eigenvalues (that is, the
{\it characteristic velocities}) and the
$\eta$-right eigenvector of the Jacobian
matrix:
\begin{eqnarray}
{\cal A}^x{ \rm _{i+{1\over 2},j,k}}
= \left(\frac{\partial \vec f(\vec u) \rm}
{\partial\vec u \rm} \right)_{{\vec u}=(
{\vec u}_{i+{1\over 2},j,k}^{L}+ {\vec u}_{i+{1\over 2},j,k}^{R})/2}
\end{eqnarray}
\noindent
calculated in the state which corresponds to the arithmetic mean of
the states at each side of the interface. Quantities
$\Delta \widetilde {\omega}_{\eta}$
-- the jumps in the local characteristic variables through each
interface-- are obtained from the following relation
\begin{eqnarray}
{\vec u}^{R} - {\vec u}^{L} = \sum_{\eta = 1}^{5}
\Delta\widetilde{\omega}_{\eta}
{\widetilde {\vec R}}_{\eta}^x
\end{eqnarray}
\noindent
where $\widetilde {\lambda}_{\eta}^x$, $\widetilde {\vec R}_{\eta}$
and $\Delta \widetilde {\omega}_{\eta}$,
are functions of $\vec u$, which
are calculated at each interface and, consequently, they depend on the
particular values
${\vec u}^{L}$ and ${\vec u}^{R}$. Analogously, numerical fluxes in
the $y$ direction, $\hat {\vec g}$, and $z$ direction , $\hat {\vec h}$,
are obtained.

\item {\it Advancing in time}:
 
Once the numerical fluxes $\hat {\vec f}$,$\hat {\vec g}$,
and $\hat {\vec h}$ are known, the evolution of quantities
${\vec u}_{i,j,k}$ is governed by
 
\begin{eqnarray}
\frac{d{\vec u}_{i,j,k}}{d t}=
-\frac{\hat{\vec f}_{i+{1\over 2},j,k}- \hat{\vec f}_{i-{1\over 2},j,k}}
{\Delta x_i}
-\frac{\hat{\vec g}_{i,j+{1\over 2},k}- \hat{\vec g}_{i,j-{1\over 2},k}}
{\Delta y_j}
-\frac{\hat{\vec h}_{i,j,k+{1\over 2}}- \hat{\vec h}_{i,j,k-{1\over 2}}}
{\Delta z_k} + {\vec s}_{i,j,k}
\end{eqnarray}
\end{enumerate}
 
An ordinary differential equation (ODE) solver derived by Shu and Osher (1988)
has been used to solve Eq (25). It is a third order Runge-Kutta that does not
increase the total variation of the numerical solution and preserves the
conservation form of the scheme.
 
The standard Courant constraint on the time step reads as follows:
\begin{eqnarray}
(\Delta t_C)_{i,j,k} = CFL_1 {\times} &{max}& \left[
 \frac{\Delta x_i}{|\lambda_1^x(x_{i+{1\over 2},j,k})
- \lambda_5^x(x_{i-{1\over 2},j,k})|}, \right. \nonumber \\
 & & \frac{\Delta y_j}{|\lambda_1^y(y_{i,j+{1\over 2},k})
- \lambda_5^y(y_{i,j-{1\over 2},k})|}, \nonumber \\
 & & \left. \frac{\Delta z_k}{|\lambda_1^z(z_{i,j,k+{1\over 2}})
- \lambda_5^z(z_{i,j,k-{1\over 2}})|} \right]
\end{eqnarray}
 
The value of the Courant time step, at each time level, is given
by $\Delta t_C = min (\Delta t_C)_{i,j,k}$, $\forall i,j,k$.
Typical values of $CFL_1$ are running from $0.6$ to $0.9$.
 
In pressureless tests, another constraint on the time step has
been included.
This is motivated by the fact that, in the absence of pressure,
Courant condition does not guarantee the
numerical stability of the system; hence,
we introduce the dynamical time
\begin{equation}
\Delta t_d=CFL_2 \sqrt{\frac{3\pi^2}{4\rho}}
\end{equation}
where $\rho$ is the maximum density
obtained in the previous time iteration.
Typical values of the factor $CFL_2$ are of the order of $10^{-3}$.
In any case, both Courant and dynamical time steps are computed
and compared. The most restrictive of these time steps is used.
 
Particular attention must be paid to the source terms. The numerical treatment
of the sources require the use of specific techniques, which depend on the
desired accuracy and the complexity of these terms. In our case, the values of
the quantities ${\vec s}_{i,j,k}$ --at each cell-- calculated at time level
$t^n$ are added to the algorithm (25) for advancing in time.

For scalar equations and in the smooth part of the flow it can be shown that
the linear reconstruction (our MUSCL-version) together with the Runge-Kutta
procedure for advancing in time (see Appendix B) set up an algorithm which is
second order accurate in space and third order accurate in time (in the sense
of the local truncation error). When the PPM's subroutine is switched on our
algorithm becomes globally third order accurate.
 
At this point several differences and analogies with the recent code published
by Ryu et al. (1993) should be pointed out. It has been shown, at a purely
theoretical level, that a MUSCL algorithm is TVD (see, e.g., LeVeque 1992).
Numerical experiments carried out in 1D Newtonian dynamics (see, e.g., Yee
1989) show that MUSCL and the TVD scheme proposed by Harten (1983) -- in which
the code developed by Ryu et al. (1993) is based on -- lead to similar results.
 
The most relevant differences with Ryu et al. (1993) code are the following:
 
1) Our code allows one to choose, by switching on an indicator, the
reconstruction --linear or parabolic-- according to the requirements between
computational cost and accuracy in each particular application. As we have
mentioned above, the parabolic reconstruction and our procedure for advancing
in time makes our code globally third order unlike the second order of Ryu et
al. (1993) code.
 
2) Our code uses a linearized Riemann solver due to Roe (1981) and when
combined with the parabolic reconstruction allows to solve the Riemann
problem with great accuracy.
 
3) Finally, let us discuss some important point concerning the so-called {\it
positively conservative} property of a HRSC scheme. In a highly energetic flow
--i.e., that one in which kinetic energy is much greater than the internal
energy-- it may happen that subtracting the kinetic energy --computed from a
conservative numerical approximation for the conservation laws of mass and
momentum-- from conservative approximation for the conservation law for the
total energy, could lead to a negative value for the internal energy, and,
hence, to a failure of the numerical scheme (Einfeldt et al, 1991). Schemes
such that both the internal energy and density remain positive throughout the
computational process have been called positively conservative by Einfeldt et
al. (1991). We do not know any theoretical analysis (theorem or lemma) proving
that Harten's algorithm (Harten, 1983) --in which Ryu's code is based on-- or
MUSCL or PPM algorithms are positively conservative. In practice, we have no
found in our numerical experiments (see below) any evidence of non-positiveness
features; in particular, the spherical shock reflection (see below) test
displays more than four orders of magnitude between kinetic energy and internal
energy without any observed failure. Ryu et al. (1993)  have overcome this
question --the non-positiveness-- by implementing in his code a bifurcation
point that, during the evolution of the system, solves a different system of
equations. In our code, if it were necessary, it is easy to implement a
modified version of Roe's Riemann solver which --via the numerical fluxes-- has
the property of being a positively conservative algorithm (Einfeldt et al.
1991).

\section{Numerical tests}
 
In this section, our multidimensional code is tested.
The most interesting aspects of each test
are pointed out. Results are discussed.
 
In all the Figures considered in this Section we use marks (crosses, circles,
squares and stars) to plot the numerical solutions obtained from our
multidimensional code, while solid lines display the features of the exact or
numerical solutions used for comparison.
 
\subsection{Tests involving shocks}
 
In the first test, we have focussed on standard {\it shock tube} problems whose
analytical solutions are well known --in terms of the initial data-- in studies
of Newtonian (see, e.g., Yee 1989) or relativistic (see Mart\'{\i} \& M\"uller
1994) computational fluid dynamics. The second test involves a spherically
symmetric shock. The exact solution can be found in Noh (1987). In these two
tests, calculations have been carried out by switching off --in our code-- the
cosmological expansion and gravity terms, and pressure obeys the equation of
state of an ideal gas with $\gamma = 5/3$.

Both tests have been done with linear and parabolic reconstruction. In general
PPM gives more accurate results. However,
the price to pay is an increasing in the
computational cost of parabolic
reconstruction, which is greater than linear reconstruction by a factor of two
or
three.

\subsubsection{Shock tube}
 
The so-called shock tube problems are a set of solutions of different Riemann
problems associated to the equations governing the dynamics of ideal gases.
They involve, in general, the presence of shocks, rarefactions and contact
discontinuities. Taking into account the fact that the analytical solution of
the Riemann problem is well-known, they are considered as standard test-beds
for checking a hydro-code.
 
In order to take advantage of the analytical solution of the standard shock
tube problem, we have considered a computational domain defined by a cube (one
unit length each edge) in which a discontinuity has been placed orthogonal to
the cube diagonal joining the points (0,0,0) to (1,1,1). This discontinuity
separates the system in two states. Initially, both states have the same
density ($\rho=1$) and velocity ($v=0$). Pressure on the left (right) hand side
of that discontinuity is $p=1$ ($p=0.1$). The grid used in these computations
has $64\times64\times64$ cells. The evolution from the above initial data leads
to the formation of a rarefaction wave,  a contact discontinuity, and a shock
travelling along the above mentioned diagonal. The location of the
discontinuity has been chosen because it leads to nonvanishing fluxes in the
three spatial directions, as occurs in the case of true 3D systems. Fig. 1
display the results obtained using linear (top panel) and parabolic (bottom
panel) reconstruction. From these panels we can see an excellent agreement
between the numerical and the analytical solutions. Let us point out that the
shock is solved in only one cell in both cases. Numerical diffusion is slightly
lesser with PPM, and the contact discontinuity is better solved with PPM as
well.

\subsubsection{Shock reflection}
 
In the spherical {\it shock reflection} test, the integration domain is a cube
initially ($t=0$) filled by a uniform cold ($p=0$) gas with density $\rho=1$.
This gas moves towards the center of symmetry with a uniform radial velocity
$v_r=-0.1$. At the central point, $r=0$, this velocity vanishes. Since the
Mach number for the infalling gas is infinite (vanishing sound velocity), the
resulting test is very severe. In Cartesian coordinates, the system can be
considered as a true 3D one. The evolution of the above initial data leads to
the formation --in the central region-- of a strong reflecting shock which
starts to propagate outwards. This shock leaves behind it a constant state of a
thermalized and compressed gas at rest. The continuous lines of Fig. 2--3
correspond to the exact solution. As it can be seen in these Figures, the shock
separates two well defined regions. More details about the exact solution and
the {\it shock reflection test} can be found in Noh (1987).

In practice we have simulated the cold gas using a specific internal energy
$\epsilon=10^{-6}$. Taken into account that in the preshock region velocity is
$v_r=-0.1$, the ratio between the thermal energy ($E_{th}$) and the total
energy (E) is $\frac{E_{th}}{E} \ge 10^{-4}$. We want to point out that even
with this strong condition the results have been acceptable, and the code seems
to support ratios between thermal and total energy up to four
orders of magnitude.
 
In the numerical experiments we discuss
in this subsection the center of symmetry is placed at
the center of a cube having an edge of 2
units length. This spatial computational domain
has been partitioned into a uniform grid of $N \times N
\times N$ cells. Two values for $N$ have been considered ($N=41,81$).
 
Figs. 2 and 3 show results obtained by
using linear and parabolic reconstruction,
respectively. They have been selected at  $t=12$.

The top (bottom) panel of Fig. 2 displays the values of the radial velocity
(density) on the diagonal joining the points (-1,-1,-1) and (1,1,1), namely, on
the main diagonal. Square and star (circle and cross) marks correspond to the
use of $41\times41\times41$ ($81\times81\times81$) cells. The squares (circles)
give the numerical values of $v_{r}$ and $\rho$ at some points of the
half-diagonal joining the points (0,0,0) and (1,1,1), while the stars (crosses)
correspond to the symmetric points with respect to the origin, which lie on the
half-diagonal joining the points (-1,-1,-1) and (0,0,0). Let us point out that
for any pair of these symmetric points, the resulting values of $v_{r}$ and
$\rho$ are indistinguishable, their relative differences  being smaller than $2
\times 10^{-5}$. It can be concluded that, for any chosen direction, the
symmetry with respect to the origin is numerically preserved by the code.
A
visual inspection of Fig. 2 shows that: (1) the shock appears located at the
position $r=r_{shock}$ given by the exact solution (continuous line), (2) it is
sharply solved in two or three cells, and (3) the most important errors appear
in the region $r<r_{shock}$. It is remarkable that these errors decrease as the
number of cells increases (compare circle with square marks); this means that
the numerical solution has a good convergence rate.
 
In the case of the experiment involving $81\times81\times81$ cells, some
quantities describing the deviations with respect to the exact solution have
been estimated. Leaving aside the points defining the shock, the maximum
relative error of density is $19 \%$ and the mean relative error is $10 \%$.
The maximum absolute error of radial velocity 
is $-6.8 \times 10^{-3}$ and the
mean absolute error is $-1.06 \times 10^{-3}$. The maximum errors appear only in
a few points located near $r=0$ and $r=r_{shock}$. Taking into account that we
are using very coarse grids, our results are encouraging. As it should happen,
the errors decrease as the number of cells increases.

Fig. 3 is analogous to Fig. 2 but
using parabolic reconstruction. The numerical solution obtained with
PPM exhibits also the fundamental property of being
symmetric in the sense explained before. Shock is solved in a
numerical cell.
The constant post-shock state in density is reached by PPM with
a resolution better than MUSCL. In particular, for the case of
poor grids ($N=41$) the differences between MUSCL and PPM are
more outstanding.

Figs. 4 and 5 display the results obtained with MUSCL at $t=15$.
 
Fig. 4 shows three slices of the spatial computational domain.
The top, intermediate and
bottom panels correspond to the coordinates $z=0.875$, $z=0$ and $z=-0.875$,
respectively.  Left (right) panel refers to densities (velocities). The top and
bottom panels of Fig. 3 show that (i) the boundary conditions on the cube faces
produce slight deviations with respect to the spherical symmetry (dark bands
near the faces), (ii) except for these small deviations, numerical solutions
show a strong spherical symmetry, (iii) the slices $z=0.875$ and $z=-0.875$ are
indistinguishable, as it is expected on account of the intrinsic features of
the code, and (iv) the shock has not arrived at these regions. Finally, the
intermediate panel shows a spherically symmetric strong shock. The region
$r<r_{shock}$ appears to be quasi-homogeneous. It only involves some small
fluctuations. The greatest deviations from the exact solution --which is
homogeneous in this region-- appear along those directions parallel to the
axis, which are contained  inside the corresponding  coordinate plane. These
effects are related to the boundary conditions.
 
Fig. 5 is presented essentially to describe the pressure behavior. The function
$p=p(t=15,x,y,z=0)$ is displayed. Let us pay attention to the following
features: The cylindrical structure of the surface (whose height is the
pressure jump at the shock), the sharp structure of the strong shock, and the
fact that boundary conditions lead to maximum errors in the directions
described above.

In our opinion, this test is so severe and the
results good enough to suggest that the use of a finer grid  --with
the corresponding consumption of machine resources (RAM memory and CPU)--
would allow to improve definitively the numerical solution.
 
In the  above numerical experiments, CPU cost -- in a HP Apollo 9000/712 --
is 0.25 ms per timestep and per numerical cell.

\subsection{Cosmological tests}
 
The third test in this paper
is based on the Zel'dovich 1D solution, which is an exact
solution of the equations describing cosmological
pressureless inhomogeneities. This case includes cosmological
expansion, but it is not a 3D case.
 
The fourth test uses a numerical planar 1D solution
of the evolution equations describing a cosmological inhomogeneity
with pressure. This solution was obtained by Quilis et al. (1994)
using a 1D code.
 
The fifth (sixth) test is based on an
analytical (numerical) spherically symmetric solution of the
equations describing the evolution of cosmological Newtonian
inhomogeneities without (with) pressure. The analytical solution
of the fifth test is described in Peebles (1980) and the
numerical solution of the sixth test was obtained by
Quilis et al. (1995).
These solutions can be considered as 1D ones
in spherical coordinates; nevertheless,
in Cartesian coordinates, they can be treated as  fully
3D solutions including cosmological expansion and gravity.

All the cosmological tests have been done with the linear reconstruction.
 
\subsubsection{Cosmological tests with planar symmetry}
 
In spite of the fact that our code is multidimensional, any 1D solution can be
used in order to perform a significant test; in fact, each of these solutions
can be used to set particular initial conditions for running our
multidimensional code. After integration (at arbitrary time), this code must
maintain the planar symmetry and reproduce the chosen planar solution. In
planar cases, Poisson's equation reduces to an ordinary differential one, which
is solved by using an ODE solver; hence, tests based on planar solutions do not
give any information about the behavior of the method used to solve Poisson's
equation (FFT) in the general case.

The {\it 1D Zel'dovich solution} is an exact solution of the
hydrodynamic equations in the planar case
(Shandarin \& Zel'dovich 1989). High density contrasts are
compatible with this solution, which is valid
before caustic formation.
Multidimensional Zel'dovich's solution is only an approximate
one. This solution is only valid in the mildly nonlinear
regime (density contrast up to a few units).
From these comments about the features of Zel'dovich's
solution, it follows that only the 1D version of this
solution is appropriate in order to test a numerical
code in the strongly nonlinear regime.
 
The first planar cosmological test is based on the 1D Zel'dovich solution.
Initial conditions are given at redshift
$Z=50$. As it is well known, Zel'dovich's
solution is completely defined by the potential of the
velocity field, $\phi_z(q)$, which is assumed to have the
following form:
$\phi_z(q)=-A\cos{kq}$, $q$ being a Lagrangian coordinate, and A and k
two free parameters. In this paper, the values of these
parameters are: $A= 2.7\times 10^{-6} h Mpc^{-1}$  and
$k=35.2$ (see Quilis et al. 1994).
The top panels of Fig. 6 show the density contrast (right)
and velocity (left) profiles at time $t/t_c=0.94$,
$t_c$ being the critical time of caustic formation for the
chosen inhomogeneity
( $t_c=1.1\times t_0$ ).
 
The results are quite good. Numerical and analytical solutions appear to be
comparable. The mean relative errors in velocity and density are of the order
of $1\%$, except at the maximum (central cell), where the relative error in the
density contrast is the greatest ( $ \sim 30\%$).

The second planar cosmological test is based on a {\it numerical 1D solution}
obtained by Quilis et al. (1994); in this
case, pressure obeys the equation of
state $p=(\gamma -1) \epsilon \rho$,
with $\gamma=5/3$; the initial value of $\epsilon$ is $10^{-6}$.
Initial conditions are given at redshift $Z=50$.
The initial profiles are the same as those in the
first cosmological planar test (see above).
In the bottom panels of Fig. 6,
the density contrast (right) and the velocity (left) profiles
are shown at time $t/t_0= 1.0$ . The solid line corresponds
to the 1D solution of Quilis et al. (1994).
In the presence of pressure,
the spatial gradients are smaller than those of the
corresponding pressureless case; for this reason,
the agreement between the two solutions compared in this test,
is greater than that of the previous planar cosmological test.
This fact is illustrated by Fig. 6, where it
can be seen that, in the presence of pressure (bottom panel),
the differences between the compared
solutions are smaller than in the pressureless case (top panel).
 
In both planar cosmological tests, the spatial grid has 200 cells in the
relevant direction.
 
\subsubsection{Cosmological tests with spherical symmetry}
 
These spherical cosmological tests have two
important features: The symmetry does not reduce
the number of relevant
directions in Cartesian coordinates, and  the general
3D Poisson equation is solved by using the FFT (Press et al. 1987).
These numerical experiments
complement previous ones because they are multidimensional
tests including gravity and expansion.
 
In the multidimensional general case, our code solves
Poisson's equation for the gravitational potential
--at each time step-- by
using the FFT. The use of this
technique requires periodic boundary conditions on
the faces of an auxiliary cube (the elemental cube
of the FFT). This fact must be carefully taken into
account in order to interpret our results.
 
The FFT is used as follows:
The density
contrast in physical space --with suitable boundary
conditions-- is the starting point. We can
distinguish three consecutive steps: (1) A FFT gives
$\delta_{\vec {k}}$ (the Fourier component of $\delta$),
(2) Poisson's equation in
Fourier space,  $\phi_{\vec {k}}=- \delta_{\vec {k}} / k^{2}$,
is used in order to get
$\phi_{\vec {k}}$, and (3)
the inverse FFT leads to the required gravitational
potential in physical space.

The first test of this section uses an {\it exact spherically symmetric
solution} of Eqs. (1)-(4) (Peebles 1980). Initial conditions are given at
redshift $Z=7$. The initial profile of the density contrast is
\begin{equation}
\delta_i(r)=\frac{\delta_c}{1+(\frac{r}{r_v})^{1.8}}
\end{equation}
with $r=\sqrt{x^2+y^2+z^2}$, $\delta_c$ being the amplitude of the density
contrast and $r_v$ the value of $r$ there 
where $\delta$ reduces to one-half of
$\delta_c$. The initial peculiar velocity corresponds to vanishing nongrowing
modes and it can be obtained from the initial density contrast. These initial
profiles are used as inputs in our code. In the applications of this paper, the
free parameters of the density profile are assumed to be $r_v=0.6 h^{-1} Mpc$
and $\delta_c=0.26$. These values were also used in Quilis et al. (1995) in
order to simulate a rich Abell cluster at redshift $Z=0.02$.
 
Although a spherical solution is used in order to give initial conditions, our
numerical code cannot reproduce this solution in all the
elemental cubes; in
order to understand this fact and interpret the results, it should be taken
into account that the FFT gives the gravitational field produced by an ideal
distribution of elemental cubes filling the space (not by a spherically
symmetric distribution of matter filling all the space). Hence, the
gravitational field is comparable to that of the spherical solution in the
central region of the cube, but not far from the
center of  symmetry, where the contribution to
the field of the neighbouring cubes is relevant.
 
From the important considerations above, it follows that: (i) our code should
reproduce the exact spherical solution in the central part of the elemental
cube and (ii) our numerical results should deviate from those of the exact
spherical solution as the distance to the symmetry center increases.
This separation should also appear in the absence of pure
numerical errors.
 
In this test, two grids  having $32 \times 32 \times 32$  and
$64 \times 64 \times 64$  cells are used and the results are compared.
The top panel of Fig. 7 displays the density contrast (right) and the
peculiar velocity (left) at time $t/t_0=0.89$. Solid line corresponds to the
exact spherical solution.
Circles  (crosses)  stand for the numerical results obtained with the
$32 \times 32 \times 32$  ($64 \times 64 \times 64$)  grid.
 In the central region of the box, the relative difference between numerical and
analytical values of $v$ and $\delta$ are smaller than $3\%$ for
the $32 \times 32 \times 32$ grid and  smaller than $1\%$ for
the $64\times 64\times 64$ grid
(except for the central point).
As expected, for both grids,
the relative differences increase as the distance to the center increases, but
they are smaller than $10\%$ everywhere. This difference
is not merely a numerical error but the superposition
of a numerical error and the deviation of two solutions coming from two
different problems: an ideal distribution of elemental cubes filling the space
and a spherically symmetric distribution of matter filling all the space.
Let us remark that satisfactory results have been obtained with the extremely
coarse grid of $32 \times 32 \times 32$ cells.

The second test of this section is based on a {\it numerical spherically
symmetric
solution} derived by Quilis et al. (1995). In this case, the initial density
profile has the form (28) and the values of
the parameters $r_v$ and $\delta_c$
are the same as in the first spherical test. The initial value
of $\epsilon$ is $5 \times 10^{-6}$.
The equation of state is $p=(\gamma -1) \epsilon \rho$, with $\gamma=5/3$.
A grid of $64 \times 64\times 64$ cells has been used.
 
The bottom panel of Fig. 7 shows the density profile (right) and the peculiar
velocity (left) at time $t/t_0=0.89$. The continuous line corresponds to the
numerical solution with pressure due to Quilis et al. (1995). Although the
initial profiles and the final time are the same as in the first spherical
test, the density contrast has reached central values smaller than in that
test. This effect is produced by the spatial pressure gradients. Apart from
this difference, both spherical tests display comparable features (see Fig.
7). Velocities show the same behavior as in the previous test.

In the above numerical calculations, CPU cost -- in a HP Apollo 9000/712 --
is 0.36 ms per timestep and per numerical cell.

\section{General discussion}

In this paper, we have numerically  solved the full
multidimensional system of hydrodynamical equations
describing the evolution of a gas --including gravity
and an expanding cosmological
background-- taking advantage of the fact that they are a system of
conservation laws with sources. This property is crucial for using
modern HRSC techniques.
 
A multidimensional hydrodynamic code based on these techniques has been
built up. This code includes the possibility of choosing between two
spatial reconstructions in order to get better resolutions, i.e. a linear
reconstruction (MUSCL) or a parabolic reconstruction (PPM).
 
An algorithm for solving Poisson's equation at
each time step is included as well.
This Poisson solver is based on the FFT. The use of this transform
is appropriated because the code described in Section. 2 gives the
density at each time step and this density is the only ingredient
required by the FFT in order to compute the gravitational potential; in other
words, if the FFT is used, no unknown boundary conditions for the
gravitational potential are required as inputs. Finally, the time elapsed by
the FFT increases very slowly as the number of points per edge --of the
elemental cube-- increases.
Although the use of the FFT has important advantages, this technique only leads
to admissible simulations in the central part of the elemental box. This is an
unavoidable limitation attached to the use of the FFT. The elemental cube must
be carefully chosen in each case.

Our code has passed successfully a battery of six severe tests. Four of them
(shock tube problem, spherical shock reflection,
Zel'dovich's solution and the Newtonian pressureless
spherically symmetric solution) are
in fact considered as standard bed-tests in
classical and cosmological hydrodynamics.
We have implemented two more numerical tests from
previous numerical solutions.
 
The behaviour of our code is good in all six cases. The tests show that the
code works in the presence of shocks, rarefactions, contact discontinuities,
cosmological expansion, gravity and pressure.
In the spherical cosmological test, it has been seen that
high density contrasts of the order of
$10^{2}$ can be reached by using spatial grids with $64 \times 64 \times 64$
points (see Fig. 7). Larger density contrasts would require a greater number
of points and, consequently, they would have a greater computational cost.
Grids having  $128 \times 128 \times 128$ nodes should allow the description of
the hot gas component located inside clusters up to density contrasts between
$10^{2}$ and $10^{3}$ (rich clusters).
 
In cosmological applications and as Ryu et al. (1993) pointed out,
it might happen that regions having large kinetical --compared with
the thermal energy-- appear. As we have discussed before, a conservative
algorithm could lead to important numerical difficulties.
In practice
we have not noticed this problem in any of the tests presented in this paper,
overall in the shock reflexion test where $\frac{E_{th}}{E} \ge 10^{-4}$
and the analytical solution was recovered quite well.
 
Important perspectives arise in the case of several cosmological problems,
specially, if the code presented in this paper is coupled to a N-body one (see
Bertschinger \& Gelb 1991 for a description of this kind of codes) describing
the evolution of the pressureless matter. In order to design this coupling, it
should be taken into account that the hot gas and the pressureless component
are gravitationally coupled. Let us point out that the resulting coupled code
could lead to very realistic simulations of rich clusters, in which, the
observed features of the baryonic component would play an important role. See
Quilis et al.(1995) for a discussion about this point.

\acknowledgements
This work has been
supported by the {\em Conselleria d'Educaci\'o i Ci\`encia de la Generalitat
Valenciana} (grant GV-2207/94) and
the  Spanish DGICYT (grant
PB94-0973). Dr. M. Steinmetz and an anonymous referee are
 acknowledged for their valuable comments and
criticism.
Authors have enjoyed fruitful conversations
with J.M$^{\underline{\mbox{a}}}$. Mart\'{\i}, J.A. Miralles, V. Romero,
J.V. Arnau and F. Banyuls.
Calculations were carried out in a
IBM 30-9021 VF at the Center de
Inform\`atica de la Universitat de Val\`encia,
and in a HP Apollo 712. V.Quilis thanks to the {\em
Conselleria d'Educaci\'o i
Ci\`encia de la Generalitat Valenciana} for a fellowship.

\appendix
\section{Spectral decompositions for the Jacobian matrices}
 
The hydrodynamic equations (5)-(7) can be written in the form of
Eq. (8), where the vector of unknowns $\vec u$ is given by Eq. (9),
and the {\it fluxes} [$\vec f, \vec g, \vec h$] in the directions
$x,y,z$ respectively,
and {\it sources} $\vec s$, are given by Eqs. (10)-(13).
 
If the Jacobian matrices in each direction,
$\cal A^x = \frac{\partial \vec f (\vec u)}{\partial \vec u}$,
$\cal A^y = \frac{\partial \vec g (\vec u)}{\partial \vec u}$, and
$\cal A^z = \frac{\partial \vec h (\vec u)}{\partial \vec u}$,
have real eigenvalues and the right eigenvectors form
a complete set then the system (8) is called an
{\it hyperbolic system of conservation laws} with sources.
 
The spectral decomposition, i.e. eigenvalues, right  and
left eigenvectors, are needed in order to build the numerical code.
For sake of simplicity we list here, the eigenvalues, and the right
 and left eigenvectors for $\cal A^x$. The spectral decompositions in
the other directions are formally identical.
 
Eigenvalues:
\begin{eqnarray}
\lambda^x_1 &=& \frac{v_x + c_s}{a} \nonumber \\
\lambda^x_2=\lambda^x_3=\lambda^x_4 &=& \frac{v_x}{a} \nonumber\\
\lambda^x_5 &=& \frac{v_x - c_s}{a}
\end{eqnarray}
where $c_s$ is the sound speed ($c_s=(\gamma p /\rho)^{1/2}$).
 
Right eigenvectors:
\begin{eqnarray}
\vec{R}^x_1 =\left( \begin{array}{c}
1\\
v_x + c_s\\
v_y \\
v_z \\
\frac{p+E}{\rho} +v_xc_s
\end{array} \right)
\, \, \, &,&
\vec{R}^x_2 =\left( \begin{array}{c}
1\\
v_x \\
v_y \\
v_z \\
\frac{v^2}{2}
\end{array} \right)
\, \, \, ,
\vec{R}^x_3 =\left( \begin{array}{c}
0\\
0\\
1\\
0\\
v_y
\end{array}\right)
\, \, \, ,
\nonumber \\
\vec{R}^x_4 &=&\left( \begin{array}{c}
0\\
0\\
0\\
1\\
v_z
\end{array} \right)
\, \, \, ,
\vec{R}^x_5 =\left( \begin{array}{c}
1\\
v_x - c_s\\
v_y \\
v_z \\
\frac{p+E}{\rho} -v_xc_s
\end{array} \right)
\end{eqnarray}
being $v^2=v_x^2+v_y^2+v_z^2$.
 
Left Eigenvectors: Orthonormal to the right vectors,
${L}_m \cdot {R}_l = \delta_{ml}$,
\begin{eqnarray*}
\vec {L}^x_1=[  \frac{v^2(\gamma -1)-2v_xc_s}{4c_s^2},
\frac{c_s-v_x(\gamma-1)}{2c_s^2}&,&-\frac{v_y(\gamma-1)}{2c_s^2},
  \nonumber \\
&-& \frac{v_z(\gamma-1)}{2c_s^2}, \frac{\gamma -1}{2c_s^2}]
\end{eqnarray*}
\begin{displaymath}
\vec {L}^x_2=[1-\frac{(\gamma -1)v^2}{2c_s^2},
\frac{v_x(\gamma-1)}{c_s^2},\frac{v_y(\gamma-1)}{c_s^2},
\frac{v_z(\gamma-1)}{c_2^2},-\frac{\gamma -1}{c_s^2}]
\end{displaymath}
\begin{displaymath}
\vec {L}^x_3=[-v_y,0,1,0,0]
\end{displaymath}
\begin{displaymath}
\vec {L}^x_4=[-v_z,0,0,1,0]
\end{displaymath}
\begin{eqnarray}
\vec {L}^x_5=[\frac{v^2(\gamma-1)+2v_xc_s}{4c_s^2},
-\frac{c_s+v_x(\gamma-1)}{2c_s^2}&,&-\frac{v_y(\gamma-1)}{2c_s^2},
\nonumber \\
&-&\frac{v_z(\gamma-1)}{2c_s^2},\frac{\gamma-1}{2c_s^2}]
\end{eqnarray}

\section{Summary of the algorithm}
 
Here, we summarize the steps in order to build up
our hydro-code. In one of them,
the spectral decompositions of the Jacobian matrices
described in the Appendix A, are needed.
For more details see  Section 2.
 
The scheme for the numerical procedure
in updating vector $\vec u^n$ to $\vec u^{n+1}$, is the following:
\begin{enumerate}
\item{}The unknowns $\vec u$ are known at the center of the numerical
cells at the time step $n$, i.e. $\vec u^n$.
\item{}Reconstruction procedure allows to compute the unknowns at
the interface between a cell and its neighbours. It must
be done for each direction. For instance, in the $x$ direction is
\begin{eqnarray}
\vec u_{i-1,j,k}^n\, ,\vec u_{i,j,k}^n\, ,\vec u_{i+1,j,k}^n
\Longrightarrow \left\{ \begin{array}{c}
{\vec u}_{i+{1\over 2},j,k}^R \\
{\vec u}_{i+{1\over 2},j,k}^L
\end{array} \right .
\end{eqnarray}
 
Reconstruction , in our code, can be linear or parabolic (PPM).
 
\item{}Numerical fluxes are computed using Roe prescription, Eq.(22). For
example in the $x$ direction,
\begin{eqnarray}
{\widehat {{\vec f}}}^n_{i+{1\over 2},j,k}
 = \frac{1}{2}\left(
( {\vec f}({\vec u}_{i+{1\over 2},j,k}^{L})^n +
{\vec f}({\vec u}_{i+{1\over 2},j,k}^{R})^n
 -\sum_{{\eta} = 1}^{5} \mid
\widetilde{\lambda}_{\eta}^x\mid
\Delta \widetilde {\omega}_{\eta}
{\widetilde {\vec R}}^x_{\eta})\right)
\end{eqnarray}
where
\begin{eqnarray}
\vec {L}^x_{\eta} \cdot ({\vec u}^{R}_{i+{1\over 2},j,k}
- {\vec u}^{L}_{i+{1\over 2},j,k} )=
\Delta\widetilde{\omega}_{\eta} \, \, \, \, \, \, , \eta=1,...,5
\end{eqnarray}
symbol ($\, \tilde{} \, $) refers to mean values in the interface.
The fluxes in directions
$y,z$ , $\vec g$ and $\vec h$, are obtained analogously.
\item{}Poisson's equation is solved by using FFT.
\item{}Sources are obtained at each cell, $\vec s (\vec u_{i,j,k})$.
\item{} Advancing in time,
\begin{eqnarray}
{\vec u}_{i,j,k}^{n+1}={\vec u}_{i,j,k}^n  - \Delta t \cal L
(\vec u^n_{i,j,k})
\end{eqnarray}
where $\Delta t=t^{n+1} - t^n$, and $\cal L$ is the operator
\begin{eqnarray}
\cal L(\vec u_{i,j,k}) &=&
\frac{\hat{\vec f}(u_{i+{1\over 2},j,k})
- \hat{\vec f}(u_{i-{1\over 2},j,k})}
{\Delta x_i}
+\frac{\hat{\vec g}(u_{i,j+{1\over 2},k})
- \hat{\vec g}(u_{i,j-{1\over 2},k})}
{\Delta y_j} \nonumber\\
&+&  \frac{\hat{\vec h}(u_{i,j,k+{1\over 2})}
- \hat{\vec h}(u_{i,j,k-{1\over 2}})}
{\Delta z_k} + {\vec s}(u_{i,j,k})
\end{eqnarray}
 
A third order Runge-Kutta,
proposed by Shu and Osher (1988), has been chosen in order to solve
Eq. (B4). The expressions corresponding to
this Runge-Kutta like solver are:
\begin{eqnarray*}
\vec u^{1}=\vec u^{n} + \Delta t \cal L(\vec u^{n})
\end{eqnarray*}
\begin{eqnarray*}
\vec u^{2}={3 \over 4}\vec u^{n} + {1\over 4}\vec u^{1}+
{1\over 4}\Delta t \cal L(\vec u^1)
\end{eqnarray*}
\begin{eqnarray}
\vec u^{n+1}={1 \over 3}\vec u^{n} + {2\over 3}\vec u^{2}+
{2\over 3}\Delta t \cal L(\vec u^{2})
\end{eqnarray}
and $\vec u^{1}$, $\vec u^{2}$ being two intermediate states.

\item{}The unknowns $\vec u$ are known at the center of the numerical
cells at the time step $n+1$, i.e. $\vec u^{n+1}$.
 
\end{enumerate}

\newpage
 
 
\figcaption{Plot of the analytical (solid line)
and two numerical (crosses)
solutions for a shock tube numerical experiment. Top 
and bottom panel
show results using linear (MUSCL) and 
parabolic (PPM) reconstruction, respectively.
In both panels, left central and right
figures display density $\rho$, pressure $p$ and  velocity
$v$, respectively, as functions of the parameter $l=(x^2+y^2+z^2)^{1/2}$.}
 
\figcaption {Plot of the analytical (solid line) and the numerical (marks)
solutions for two reflection shock experiments at $t=12$ using
linear reconstruction (MUSCL).
Top (bottom) panel
corresponds to radial velocity (density). Squares (stars)
represent points of the half-diagonal joining the points
(0,0,0) and (1,1,1) ((-1,-1,-1) and (0,0,0)) in the experiment
involving $41 \times 41 \times 41$ cells. Circles and crosses play
the role of  squares and stars, respectively, in the case of the
experiment using $81 \times 81 \times 81$ cells.}

\figcaption {Analogous to Fig. 2 but using parabolic reconstruction
(PPM).}
 
\figcaption {Left (right) panels show density (radial
velocity) at $t=15$. Top, intermediate and bottom panels correspond to
$z=0.875$, $z=0$, and $z=-0.875$, respectively. At the left side
of each panel, it is placed a palette displaying the grey scale.}
 
\figcaption {Plot of the function $p=p(t,x,y,z)$ for $z=0$ and $t=15$.}
 
\figcaption {Left (right) top panel shows the peculiar velocity (density
contrast) for the Zel'dovich solution (solid line) and the numerical one
(crosses) at time $t/t_0 = 1$. Similar for the bottom panels, where the solid
line corresponds to a numerical solution due to Quilis et al. (1994).}
 
\figcaption {Left (right) top panel shows the peculiar velocity (density
contrast) for a pressureless spherical solution (solid line) and the numerical
one (circles for $32 \times 32 \times 32$  and crosses for
$64\times 64\times 64$ cells)
at time $t/t_0 = 0.89$. Similar for the bottom panels, where the
solid line corresponds to a numerical spherically symmetric solution due to
Quilis et al. (1995).}
 
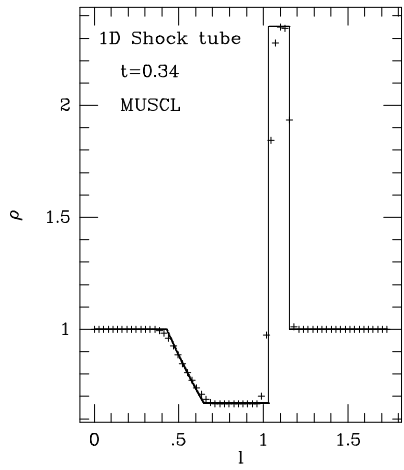
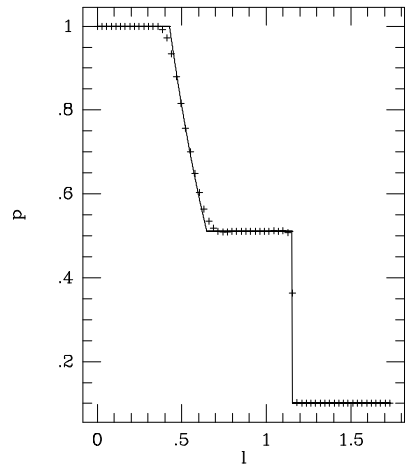
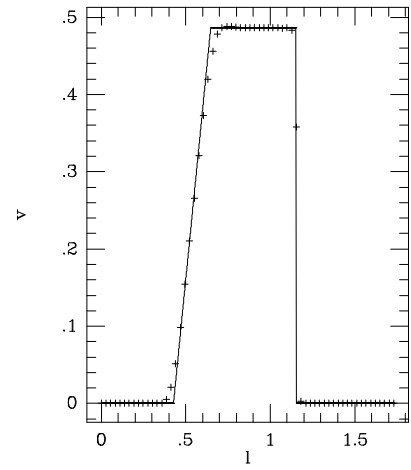
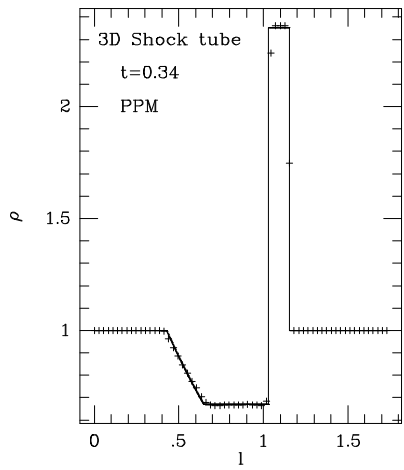
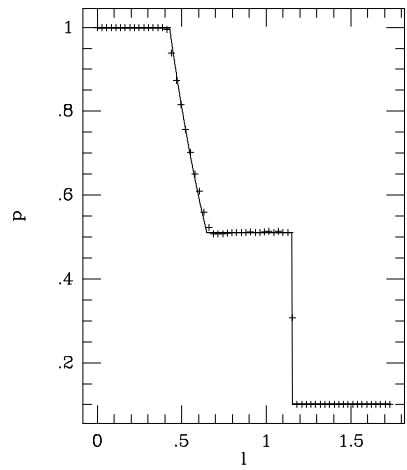
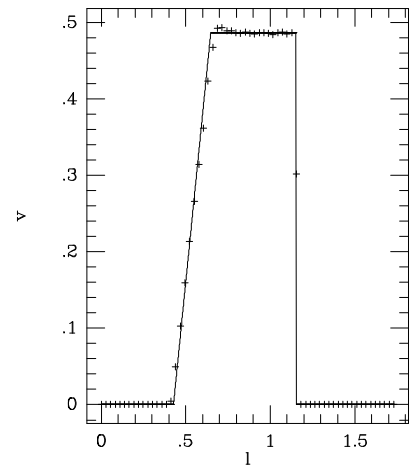
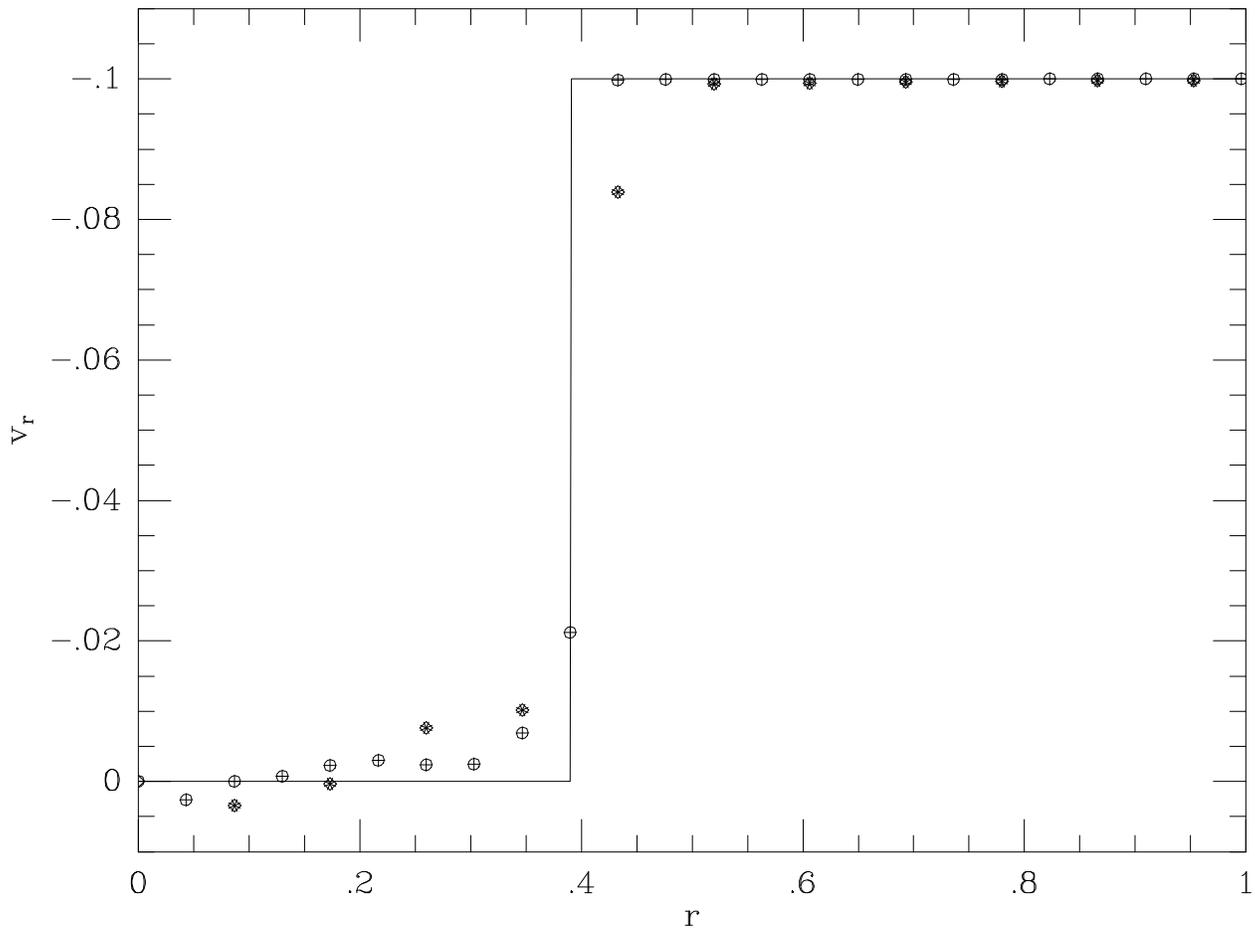
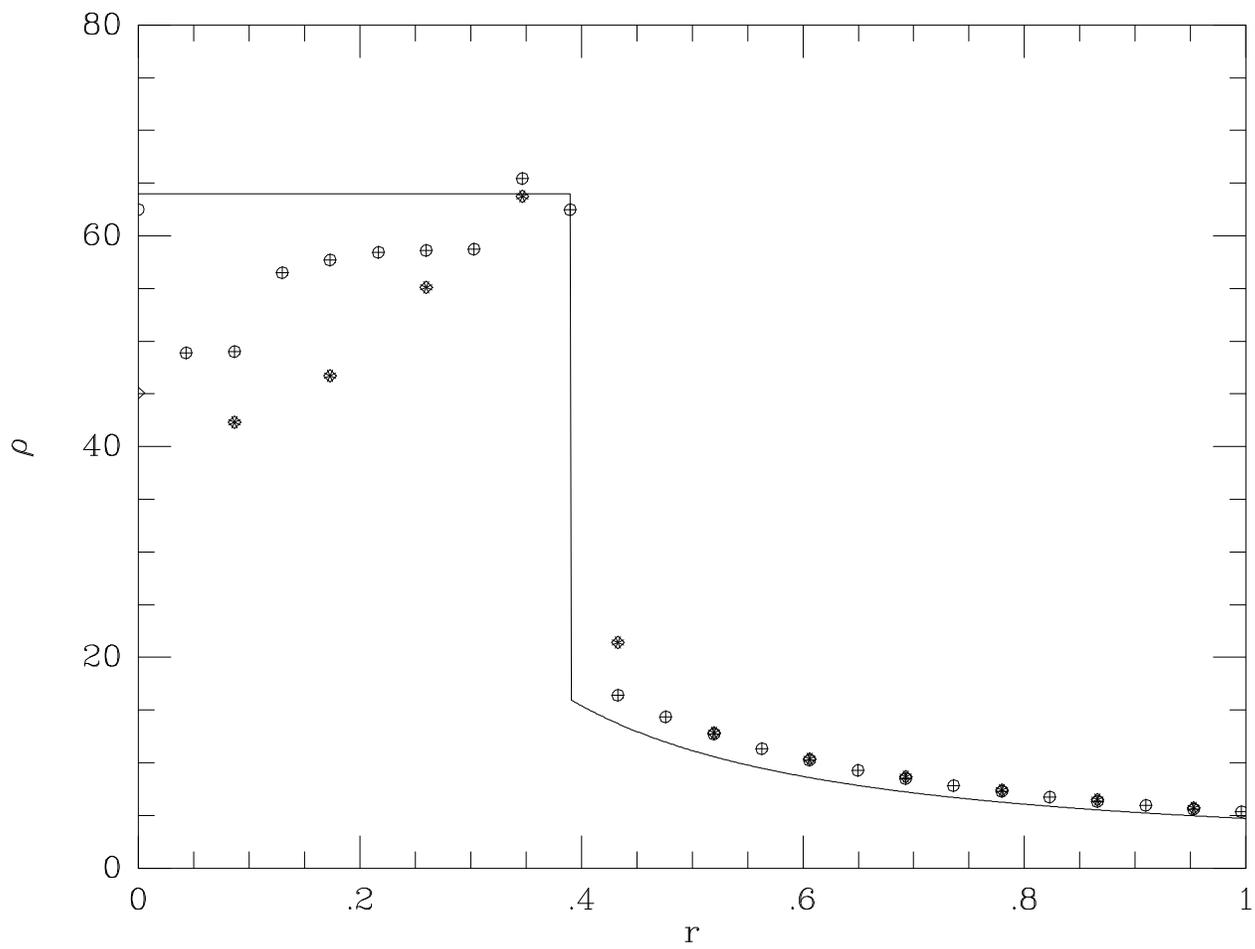
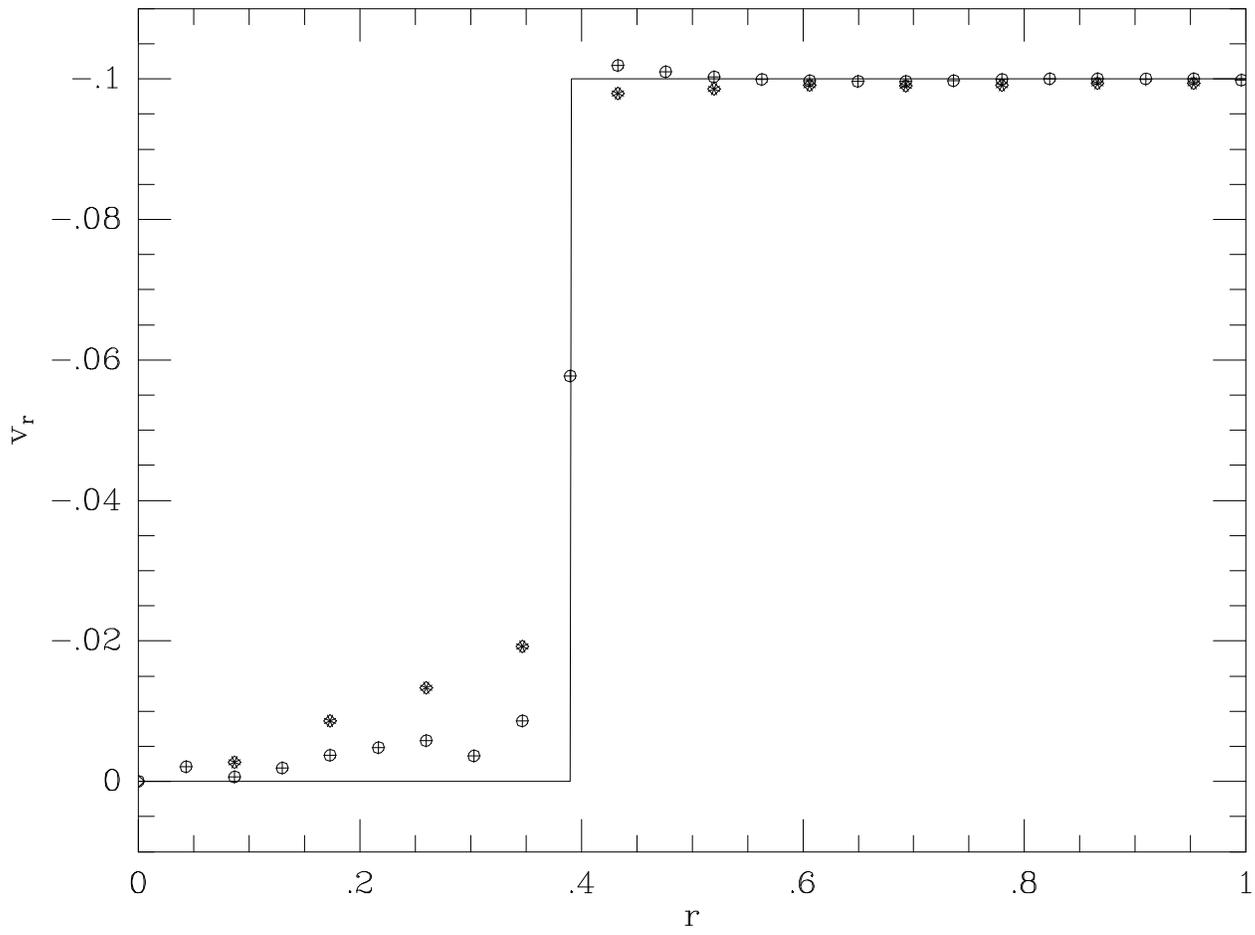
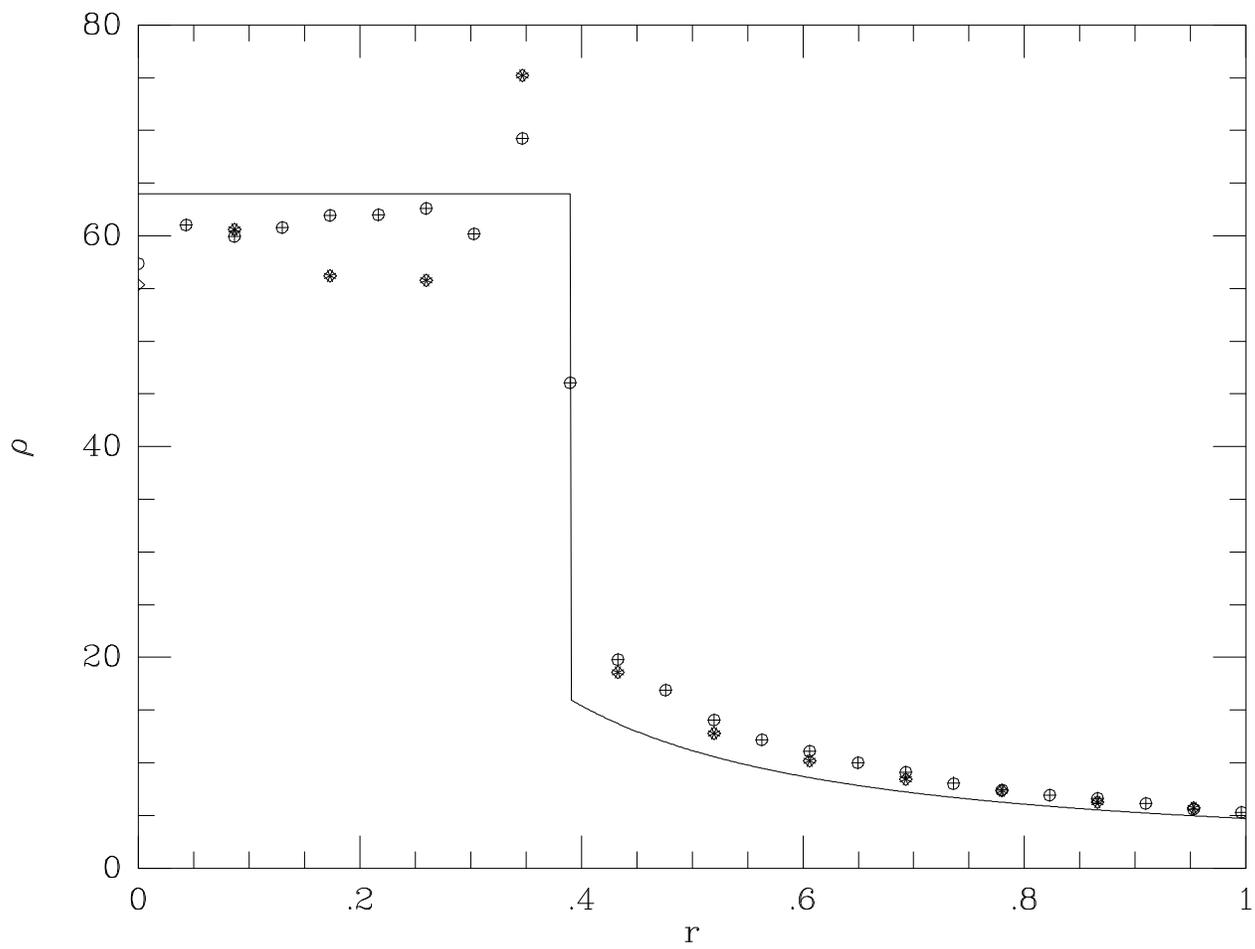
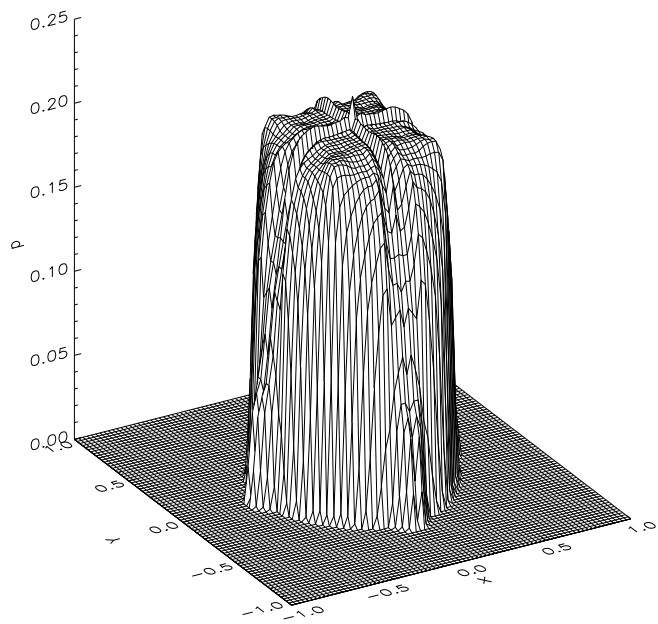
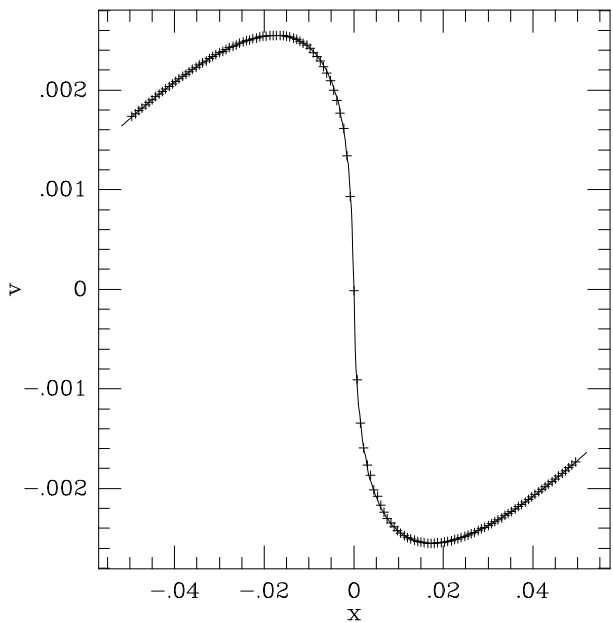
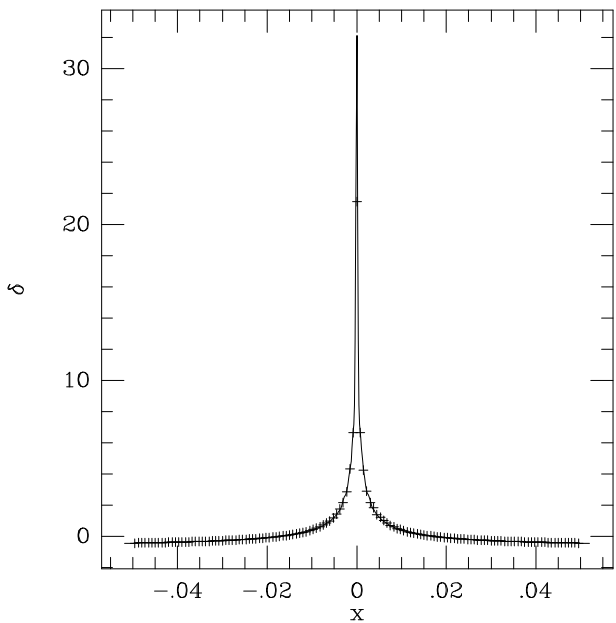
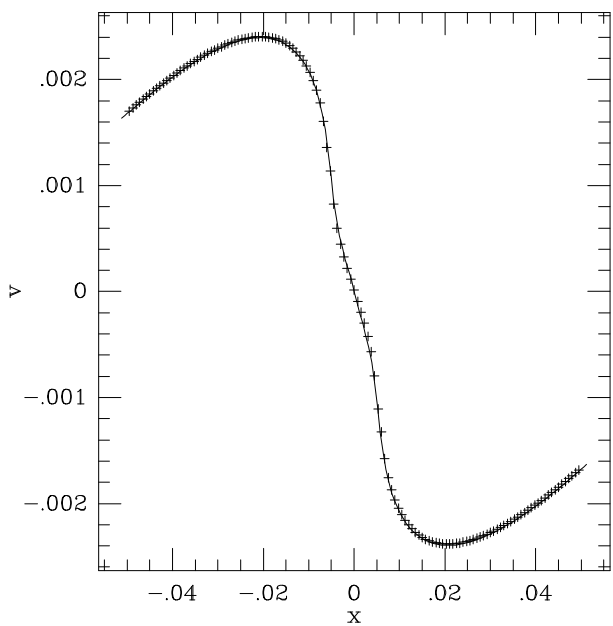
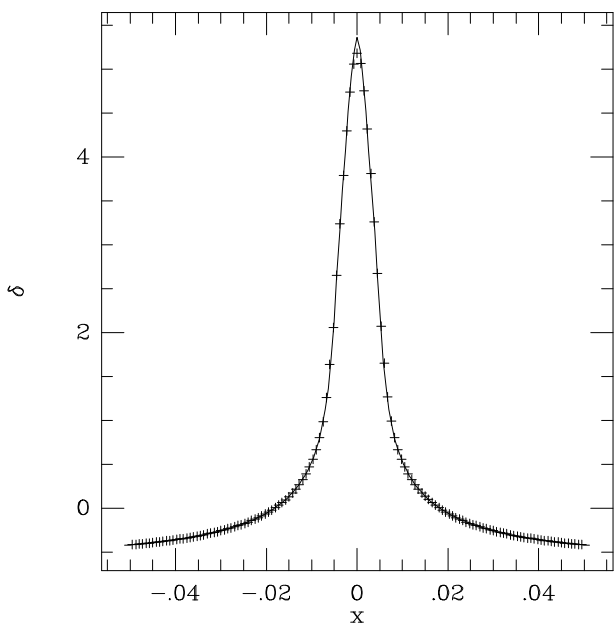
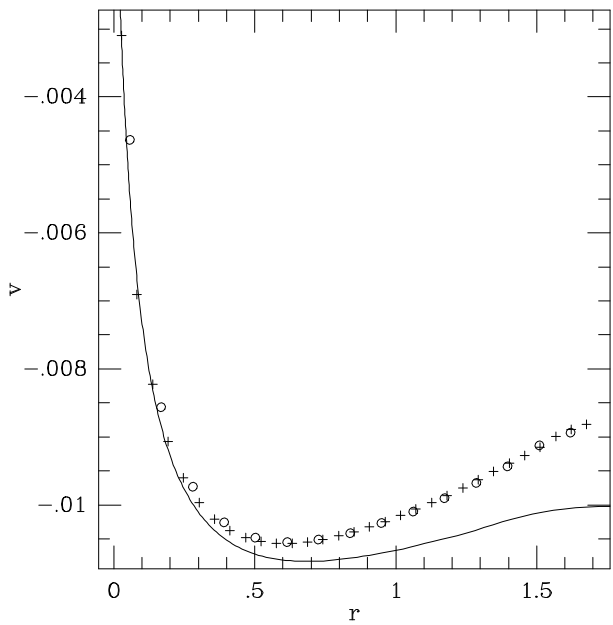
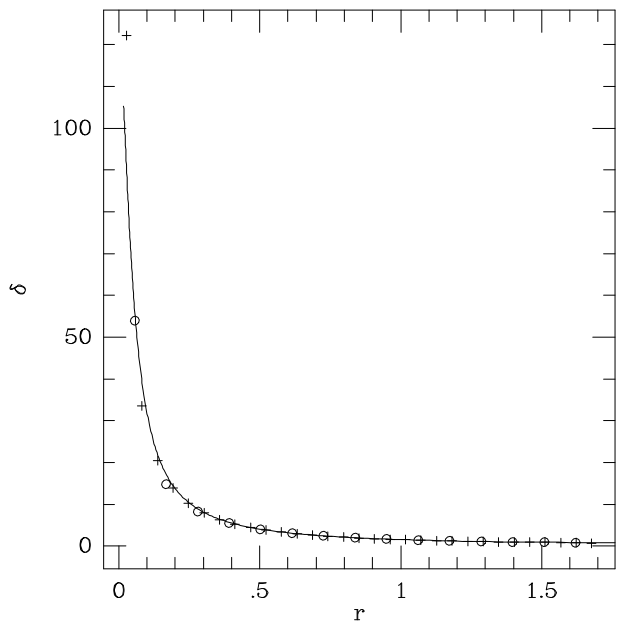
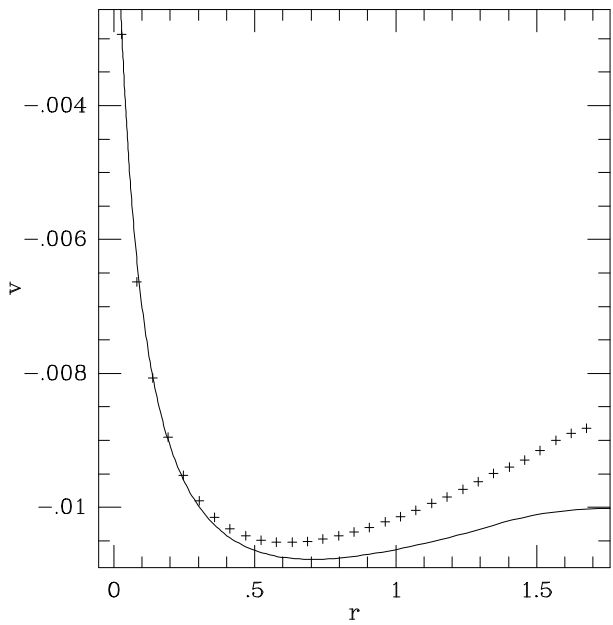
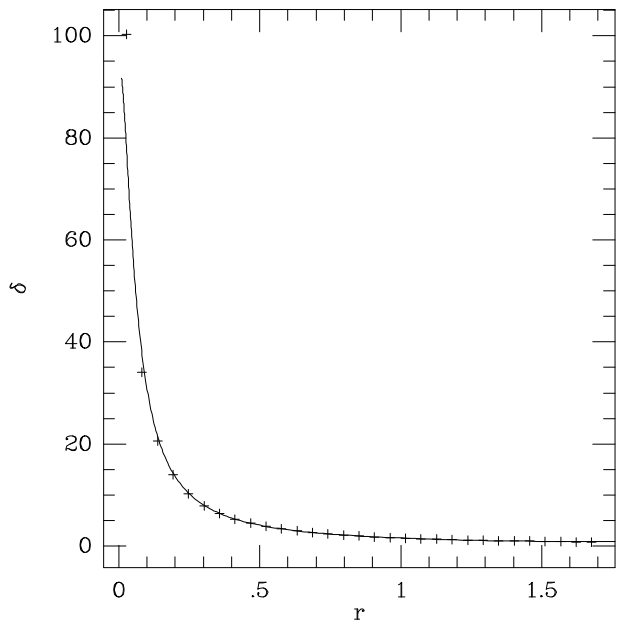
\end{document}